\documentclass[article]{aa}  

\usepackage{graphicx}
\usepackage{epstopdf}
\usepackage{txfonts}
\usepackage{color}

\begin{document}

   \title{OSSOS: XV. No active Centaurs in the Outer Solar System Origins Survey}

   \author{
Nahuel Cabral\inst{1}
\and
Aur\'elie Guilbert-Lepoutre\inst{1}
\and
Wesley C. Fraser\inst{2}
\and
Micha\"el Marsset\inst{2}
\and 
Kathryn Volk\inst{3}
\and
Jean-Marc Petit\inst{1}
\and
Philippe Rousselot\inst{1}
\and
Mike Alexandersen\inst{4}
\and
Michele T. Bannister\inst{2}
\and
Ying-Tung Chen\inst{4}%(\Chi{"96}{"73}\Chi{"82}{"F1}\Chi{"54}{"0C})
\and
Brett Gladman\inst{5}
\and
Stephen D. J. Gwyn\inst{6}
\and
J. J. Kavelaars\inst{6,7}
          }

\institute{
Institut UTINAM, UMR6213 / CNRS - UBFC, 25000 Besancon, France
\email{nahuel.cabral@utinam.cnrs.fr}
\and              
Astrophysics Research Centre, School of Mathematics and Physics, Queens University Belfast, Belfast BT7 1NN, United Kingdom
\and
Lunar and Planetary Laboratory, University of Arizona, Tucson, AZ 85721, USA
\and
Institute of Astronomy and Astrophysics, Academia Sinica; 11F of AS/NTU Astronomy-Mathematics Building, Nr. 1 Roosevelt Rd., Sec. 4, Taipei 10617, Taiwan
\and
University of British Columbia, 6224 Agricultural Road, Vancouver, BC V6T 1Z1, Canada
\and
Herzberg Astronomy and Astrophysics Research Centre, National Research Council of Canada, 5071 West Saanich Rd, Victoria, British Columbia V9E 2E7, Canada
\and
Department of Physics and Astronomy, University of Victoria, Elliott Building, 3800 Finnerty Rd, Victoria, BC V8P 5C2, Canada
}

   \date{Received August 6, 2018; accepted September 24, 2018}

  \abstract
   {Centaurs are icy objects in transition between the transneptunian region and the inner solar system, orbiting the Sun in the giant planet region. Some Centaurs display cometary activity, which cannot be sustained by the sublimation of water ice in this part of the solar system, and has been hypothesized to be due to the crystallization of amorphous water ice.}
   {In this work, we look at Centaurs discovered by the Outer Solar System Origins Survey (OSSOS) and search for cometary activity. Tentative detections would improve understanding of the origins of activity among these objects.}
   {We search for comae and structures by fitting and subtracting both Point Spread Functions (PSF) and Trailed point-Spread Functions (TSF) from the OSSOS images of each Centaur. When available, Col-OSSOS images were used to search for comae too.}
   {No cometary activity is detected in the OSSOS sample. We track the recent orbital evolution of each new Centaur to confirm that none would actually be predicted to be active, and we provide size estimates for the objects.}
   {The addition of 20 OSSOS objects to the population of $\sim$250 known Centaurs is consistent with the currently understood scenario, in which drastic drops in perihelion distance induce changes in the thermal balance prone to trigger cometary activity in the giant planet region.}


   \maketitle
   
%Lawler
%==========================================================
\section{Introduction}
Centaurs are a population of icy objects orbiting the Sun in the giant planet region: they are typically defined as having a perihelion distance between 5.2 and 30~au, and a semi-major axis below 30~au \citep{Jewitt2009}.
While most Centaurs come from the Scattered Disk \citep{DuncanLevison1997,VolkMalhotra2008}, \citet{DiSisto2010} showed that a small fraction may come from the Plutino population, and \citet{HornerLykawka2010} suggested that some could possibly come from the Neptune Trojan clouds. \citet{Lawler2018} measured the intrinsic Centaur population as 130$^{+80}_{-60}$ objects with H$_r$<8.66, and 3700$^{+2100}_{-1600}$ objects with H$_r$<12 , with 95\% confidence. The orbital evolution of Centaurs is strongly influenced by close encounters with the giant planets. In particular, the mean dynamical lifetime of Centaurs ranges from 10 to 100~Myr \citep{LevisonDuncan1997, TiscarenoMalhotra2003, Horner2004, DiSistoBrunini2007}.

Within the known Centaur population, 10 to 20\% of objects display cometary activity. The typical driver for activity is sublimation, especially sublimation of water ice when comet nuclei get closer than 3~au to the Sun. In the giant planet region however, water ice is thermodynamically stable: it does not sublimate, and thus cannot drive the activity of Centaurs. 
While trying to identify the origin of Centaurs' activity, \citet{Jewitt2009} noticed that the physical and orbital properties of active Centaurs are such that their activity should be thermally driven. \citet{GuilbertLepoutre2012} studied the phase transition between amorphous and crystalline water ice, an irreversible process releasing trapped volatile molecules \citep{BarNun1985, Laufer1987, HudsonDonn1991, JenniskensBlake1994, NotescoBarNun1996, BarNunOwen1998, Notesco2003}, as a possible source for Centaurs activity. Their main main findings are as follows:
\begin{itemize}
\item Crystallization can indeed be a source of activity in the giant planet region, possibly triggered at heliocentric distances as large as 16~au. 
\item Due to the release of trapped volatiles during the phase transition, crystallization is an efficient source of outgassing for heliocentric distances up to 10-12~au.
\item Due to the propagation of the crystallization front below the surface, crystallization-driven activity would only be sustained for a limited time, typically hundreds to thousands of years. In the best case scenarios, it could be sustained for up to tens to hundreds of thousand years. 
\end{itemize}
These findings imply that if crystallization is indeed the driver for the activity of Centaurs, these Centaurs should have suffered from a recent orbital change. This conclusion obtained from thermal evolution modeling seems confirmed by the work by \citet{Fernandez2018}, who studied the dynamical evolution of both active and inactive Centaur subpopulations. They showed that while all Centaurs may come from the same source region, they have a wide range of dynamical histories, and do not contribute to the same comet populations. Active Centaurs were found to have experienced drastic drops in their perihelion distances in the recent past. The timescales -- 10$^2$ to 10$^3$ yrs -- is compatible with the results from \citet{GuilbertLepoutre2012}. After a drastic change in orbital parameters, a Centaur would need to adjust to new thermal conditions, at which point phase transitions could be triggered and produce some observable cometary activity. No such drops were found for the inactive subpopulation, except for one Centaur --2003 CC$_{22}$-- which has so far remained inactive \citep{Fernandez2018}. We note that its perihelion distance dropped from 5.26 to 4.16~au, so it is very possible that the fuel for cometary activity was already exhausted before the drop as the object was already close enough to the Sun to be active. 

In this work, we search for cometary activity among Centaurs recently detected by the Outer Solar System Origins Survey (OSSOS). The coma search reveals that none of these objects are active within our detection limits, though a few of them do have orbits which would allow them to be active if they had suffered from a recent orbital change. We then track their past orbital evolution to test if this can explain the Centaurs' lack of noticeable activity. Finally, we provide estimates for the sizes of these Centaurs.

%
%==========================================================
\section{Centaurs in the OSSOS sample}
\begin{figure}
\includegraphics[width=\hsize]{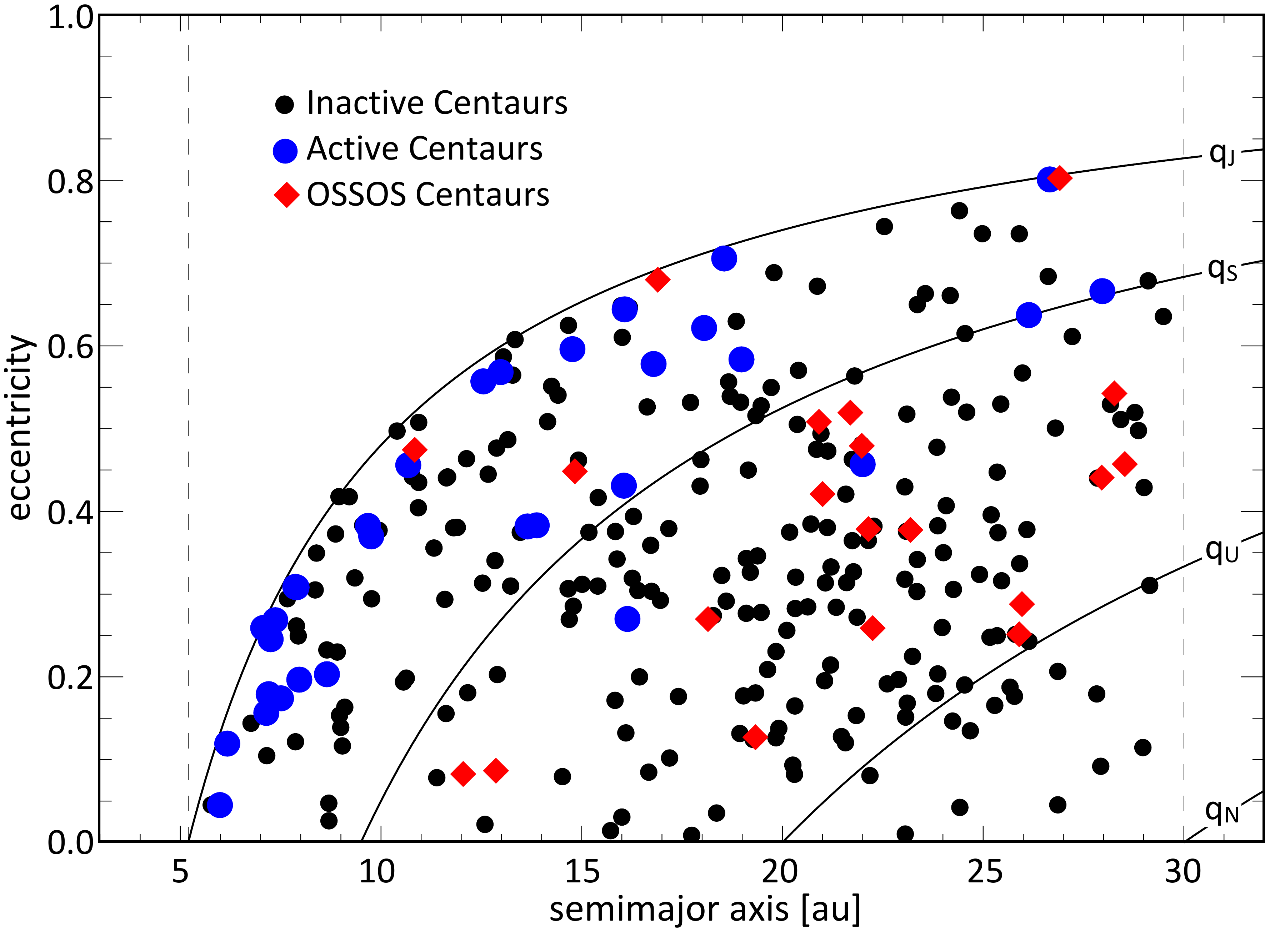}
      \caption{Semi-major axis vs. eccentricity of the orbits of Centaurs as defined with 5.2$<q<$30 au and $a<30$au. Inactive and active Centaurs are shown with small black dots and large blue dots respectively. Centaurs detected by OSSOS are shown with red diamonds. Solid curves show the orbits having perihelia distances equal to the semi-major axis of Jupiter (q$_J$), Saturn (q$_S)$, Uranus (q$_U$) and Neptune (q$_N$). Dashed lines show semimajor axis limits of 5.2 and 30au.}
         \label{orbits}
\end{figure}

OSSOS is a Large Program using the wide-field imaging instrument MegaPrime on the 3.6m Canada-France-Hawaii Telescope \citep[][PI: B. Gladman]{Bannister2016}. It surveyed 155 deg$^2$ of the sky in r-band, down to a depth of 24.1 to 25.2, thus detecting 840 new objects in the outer Solar System \citep{Bannister2018}. 
In this work, we use the simple but practical definition of Centaurs proposed by \citet{Jewitt2009}. Those are defined as objects whose orbits satisfy the following criteria: 
\begin{itemize}
\item their perihelion distance $q$ must be between 5.2 and 30~au, 
\item their semi-major axis $a$ must be below 30~au,
\item they must not be in a 1:1 long-term stable mean-motion resonance with a planet (i.e. we exclude in particular Jupiter and Neptune Trojans).
\end{itemize}
Using these constraints in the JPL small-body database search engine \footnote{https://ssd.jpl.nasa.gov/sbdb$\_$query.cgi} yields a total of 226 inactive objects and 31 active objects. These are displayed in Figure \ref{orbits}, which shows the distribution of Centaurs in the semi-major axis vs. eccentricity plane. We use the same criteria for selecting Centaurs in the OSSOS detections \citep{Bannister2018}, which yields additional Centaurs. Their orbital parameters are given in Table \ref{orbparam}. The perihelion distance of 4 of these OSSOS Centaurs is below 10~au, where cometary activity is prone to be initiated, and 5 additional objects have a perihelion distance between 10 and 12~au, i.e. in the region where cometary activity is still possible \citep{GuilbertLepoutre2012}. We note that two of these objects were not tracked (suffix $nt$ in Table \ref{orbits}), and thus have large uncertainties in their orbital elements \citep[see Sec. 2.9.1 in][]{Bannister2018}. For instance, the uncertainty in their semi-major axis is of the order of 50\%, which means in a sense that they should not appear in the table for orbital elements in a first place. However, they were discovered at a distance where we can still be sure that they are Centaurs (the distance has an uncertainty of about 20\%), so that it is acceptable to comment on their potential activity. For comparison, the semi-major axis uncertainty of tracked objects is typically 0.006~au or smaller, except for K13U17U which has an uncertainty of 0.03~au.

\begin{table}
\caption{Osculating barycentric orbital elements of Centaurs detected by OSSOS} 
\label{orbparam}
\centering
\begin{tabular}{rlllrrr}
\hline\hline 
 \# & Name & ~$a$ [au] & ~~~~$e$ & $q$ [au]~ & $i$ [deg] & $\text{a}_{eq}$ [au]\\ 
\hline                      
  1 & K13J64C    & 22.144 & 0.378 & 13.761 & 32.021 & 18.980\\%o3o01
  2 & K13U17C    & 19.328 & 0.127 & 16.872 & 32.476 & 19.016\\%o3l02
  3 & K13U17U    & 25.899 & 0.251 & 19.387 & 8.516  & 24.267\\%o3l03
  4 & K14UM5J    & 23.189 & 0.377 & 14.433 & 21.319 & 19.893\\%o4h01
  5 & K14UM9G    & 27.953 & 0.440 & 15.631 & 12.242 & 22.541\\%o4h02
\hline  
  6 & K15G53Y   & 12.048 & 0.082 & 11.052 & 24.112  & 11.967\\%o5p001
  7 & o5p002nt  & 14     & 0.4   & 8      & 6       & 11.760\\
  8 & K15H09P   & 18.145 & 0.269 & 13.248 & 3.070   & 16.832\\%o5p003
  9 & K15G54B   & 21.000 & 0.420 & 12.161 & 1.628   & 17.296\\%o5p004
 10 & K15G54A   & 22.247 & 0.258 & 16.488 & 11.401  & 20.766\\%o5p005
\hline  
11 & K15KH2H    & 16.894 & 0.679 & 5.406  & 9.083   & 9.105\\%o5m01
12 & K15KH2J    & 10.841 & 0.474 & 5.698  & 11.403  & 8.405\\%o5m02
13 & K15RR7K    & 26.910 & 0.802 & 5.308  & 9.533   & 9.601\\%o5s01
14 & o5s03nt    & 13     & 0.1   & 12     & 90      & 12.870\\
15 & K15RR7H    & 20.910 & 0.508 & 10.284 & 10.110  & 15.514\\%o5s04
\hline  
16 & K15RO5V    & 21.976 & 0.479 & 11.446 & 15.389  & 16.934\\%o5s05
17 & K15RR7F    & 21.692 & 0.519 & 10.426 & 0.927   & 15.849\\%o5t02
18 & K15RR7D    & 25.967 & 0.288 & 18.488 & 18.849  & 23.813\\%o5t03
19 & K15VG4E    & 28.529 & 0.457 & 15.487 & 36.539  & 22.571\\%o5c001
20 & K15VG4F    & 28.271 & 0.542 & 12.933 & 5.729   & 19.966\\%o5d001

\hline                           
\end{tabular}
\end{table}

%==========================================================
\section{Coma search}
We started by visually inspecting the images in order to look for any obvious coma, which we did not find, comparing images taken over the duration of the OSSOS program (of the order of 15-20 images per object, taken both in $r-$ and $w-$ bands, see \citet{Bannister2018} for details). We then determined the brightness profile of each target, from which we can search for cometary activity by comparing it to the Point-Spread Function (PSF) radial profile computed from stars in the field. This procedure was applied independently, using two different methods, and yielded the same result: no cometary activity can be detected on Centaurs in the OSSOS sample.

\begin{figure}
\includegraphics[width=\hsize]{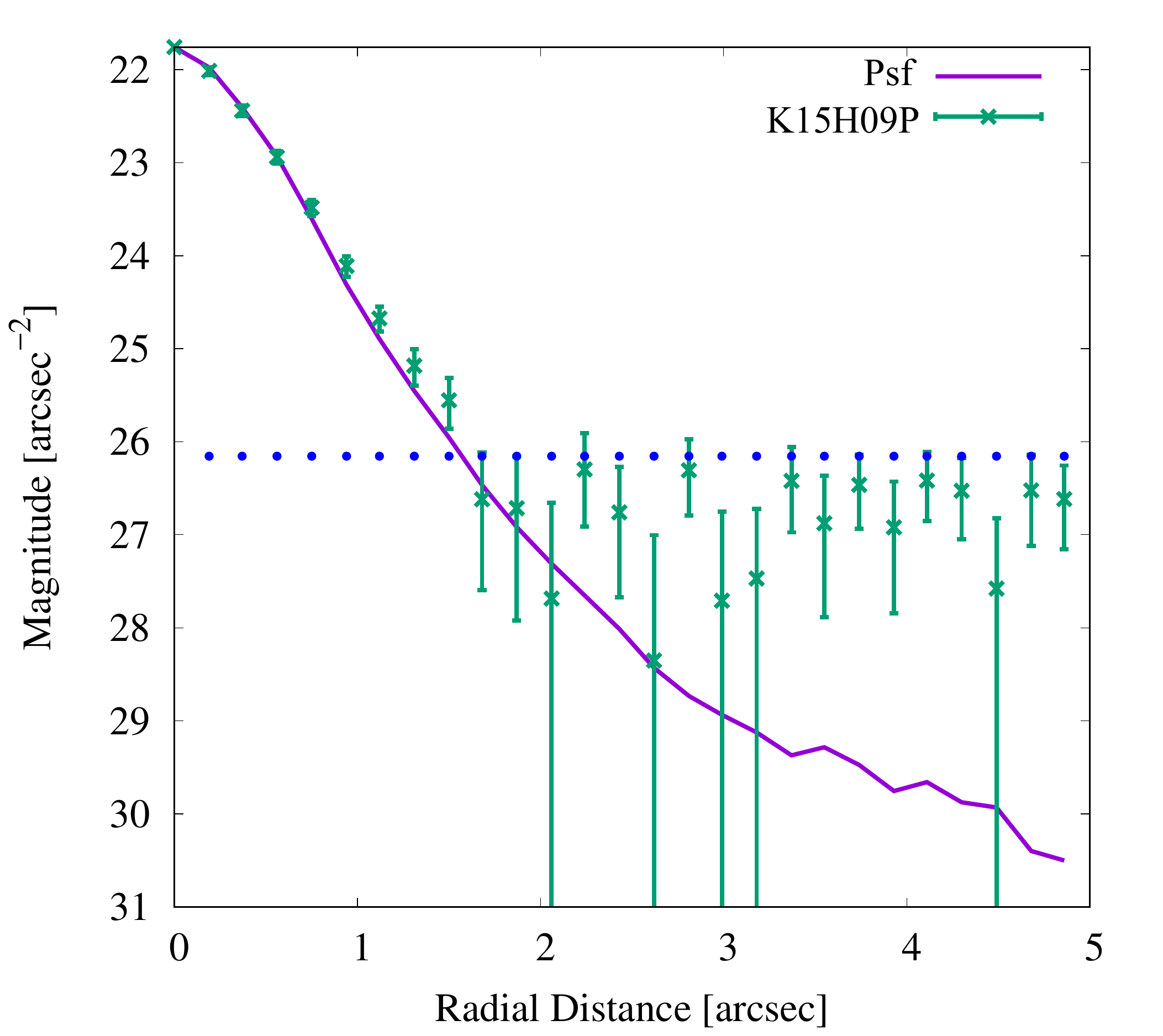}
  \caption{Surface brightness profile of Centaur K15H09P (OSSOS image) compared to the average PSF radial profile from stars in the field. The dotted line shows the surface brightness of the background sky.}
         \label{profile}
\end{figure}

For the first method, we used the standard IRAF functions in the noao.digiphot.daophot package\footnote{http://iraf.noao.edu/} to determine the targets centroid, and sExtractor \footnote{https://www.astromatic.net/software/sextractor \citep{BertinArnouts1996}.} to subtract the sky background and produce the radial brightness profiles. No object was found to be broader than the profile of stars (Figure \ref{profile}). To add a degree of accuracy, we then used the Trippy software\footnote{https://github.com/fraserw/TRIPPy} developed specifically for moving targets by \citet{Fraser2016} and trailed images. With this software, we can calculate, fit and subtract a Trailed point-Spread Functions (TSF) for each OSSOS image. We could not find any evidence for a coma either. 
The residual do not show any structure and fluxes are comparable to the background flux, or to the residual of PSF-subtracted stars of similar magnitudes when available. However, it is worth noting that most objects are only slightly brighter than the detection limits: this means that we are only sensitive to high levels of activity. Table \ref{size} gives $f_{lim}$, the minimum fraction of coma flux to which OSSOS images are sensitive, based on detection-limit magnitudes. Only 4 objects are sufficiently bright to allow the detection of comae which could contribute to 30\% or less of the total flux. Therefore, our data do not allow to completely rule out any activity, as these objects may be active below out detection thresholds, as seen in the early ESA/Rosetta data \citep{Snodgrass2016}.

\begin{figure}
\includegraphics[width=\hsize]{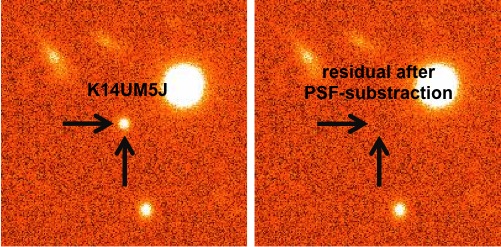}
      \caption{Left: Col-OSSOS image of Centaur K14UM5J. Right: residual image after the object has been fitted with a PSF which is then subtracted. No remaining structure can be seen in the residuals.}
         \label{psf}
\end{figure}

Among the 4 OSSOS Centaurs with enough signal to noise to reach low levels of activity, Col-OSSOS data (which are deeper than the OSSOS data) exist for K14UM5J, K15RR7K, and K15RO5V. Col-OSSOS is a large program \citep[][Schwamb et al. subm.\footnote{https://arxiv.org/abs/1809.08501}, PI: W. Fraser]{Pike2017} with the 8.1-m Frederick C. Gillett Gemini North Telescope also on Mauna Kea, which that collects near-simultaneous \emph{g}, \emph{r}, and \emph{J} band photometry of a magnitude-limited (\emph{r}$<$23.6) subset of the OSSOS sample. Optical measurements were acquired with the Gemini Multi-Object Spectrograph (GMOS, \citealt{Hook}), and the \emph{J} band sequence was obtained with the Near InfraRed Imager and Spectrometer (NIRI, \citealt{Hodapp}). Even with these deeper images, no cometary activity was found either by PSF fitting or inspecting the residuals from the PSF subtraction (Fig.\ref{psf}).

%==========================================================
\section{Dynamical evolution}
As stated above, \citet{GuilbertLepoutre2012} found that if crystallization of amorphous ice is the source of Centaurs' cometary activity, it could only be sustained for a limited time. In this scenario, the emergence of activity would require a drastic change in active Centaurs' orbital parameters in a recent past, typically of the order of 10$^{2-3}$ years. This timescale might be increased to 10$^{4-5}$ years for some best case scenarii, i.e. for certain combinations of the thermal conductivity, albedo and orbital obliquity (parameters tested in their simulations) which may not be representative of the majority of actual Centaurs. We have thus tracked the past dynamical evolution of the OSSOS Centaurs (though not for the 2 $nt$ objects) in order to assess whether any of them would have suffered from a drastic orbital change able to induce cometary activity. To do so, we have integrated the orbit of each OSSOS Centaur backwards for 10$^5$ years with clones showing the best-fit orbit, and the 3$\sigma$ minimum and maximum semimajor axis orbits allowed by the astrometry. The integrations are stopped when the clone reaches a heliocentric distance smaller than 4 au because the simulations only include the Sun and the 4 giant planets as massive perturbers (the mass of the terrestrial planets is added to the Sun). The simulations were performed using the rmvs3 routine in the SWIFT integrator package \citep{LevisonDuncan1994} with a timestep of 30 days.

\begin{figure}
\includegraphics[width=\hsize]{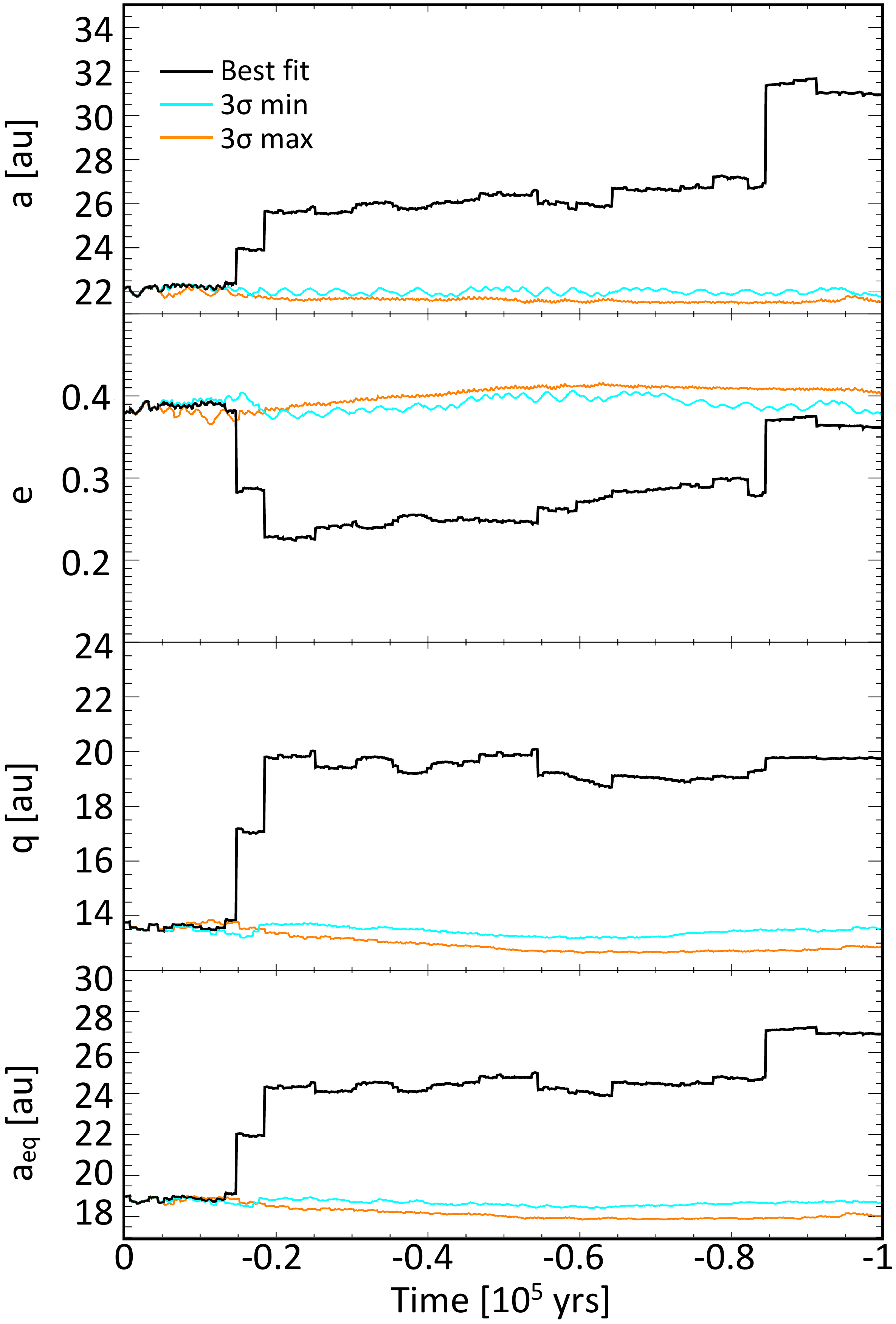}
      \caption{From top to bottom: evolution of the semimajor axis, eccentricity, perihelion distance and equivalent circular orbit of K13J64C
      %o3o01
      for the past 10$^5$ years. The black curves show the evolution of the best-fit orbit, while the colored curves show the evolution of orbits with semimajor axes approximately 3$\sigma$ above and below the best fit orbit.}
         \label{o3o01}
\end{figure}

Some OSSOS Centaurs evolved smoothly in the past 10$^5$ years, at distances in the giant planet region where cometary activity would not be expected from thermal evolution models. We also found objects suffering from drastic changes in orbital parameters, as shown in Figure \ref{o3o01}. In these plots, we follow the changes in semimajor axis ($a$), eccentricity ($e$), perihelion distance ($q$) and equivalent distance ($a_{eq}$) for the past 10$^5$years. The later, a$_{eq}$=$a(1-e^2)$, corresponds to the semimajor axis of a circular orbit receiving the same amount of energy from the Sun as the real orbit. It allows direct comparisons with simulation results from \citet{GuilbertLepoutre2012}. The best-fit orbit for K13J64C
%o3o01
had a drop in its perihelion distance of more than 6~au about 2$\times$10$^4$years ago. This translates into a drop of about 4~au in $a_{eq}$. Even if in theory such a drop would be able to trigger cometary activity, a- it happened too long ago for effects to be still seen today, and b- the object remains much too far from the Sun for crystallization to be triggered. 

\begin{figure}
\includegraphics[trim={0 0 0 0cm},clip,width=\hsize]{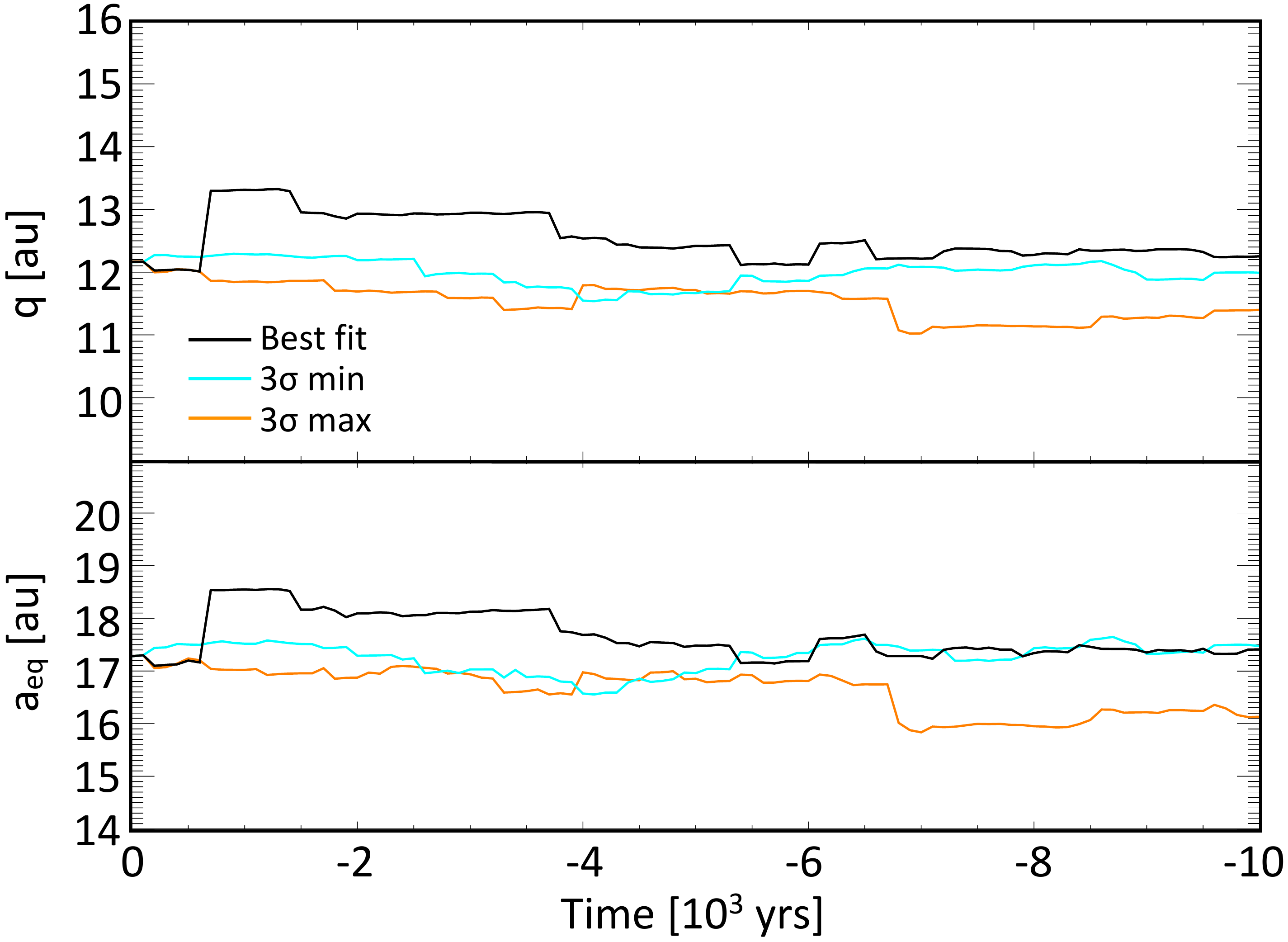}
      \caption{From top to bottom: evolution of the semimajor axis, eccentricity, perihelion distance and equivalent circular orbit of K15G54B for the past 10$^4$ years.}
         \label{o5p004}
   \end{figure}

We searched for more moderate drops in the past 10$^{3-4}$ years, as shown in Figure \ref{o5p004}. The object K15G54B %o5p004 
suffered from a $\sim$1.5~au drop in both perihelion and equivalent distances about 700~years ago. However, the object remains too far from the Sun, with an equivalent distance of about 17~au (where no activity due to crystallization can be triggered, and we also note that $q$ remains beyond 12~au). The results of our dynamical evolution analysis show that OSSOS Centaurs either did not suffer from any orbital change bringing them significantly closer to the Sun, or they did but remained too far from the Sun for activity to be triggered. The objects K15KH2H, K15KH2J and K15RR7K, %o5m01, o5m02 and o5s01, 
with perihelion distances close to Jupiter, have been on orbits close to the Sun for at least the past 10$^5$years, and are therefore not expected to be active today. 

Was OSSOS particularly unlucky in finding only Centaurs relatively stable on a timescale of 10$^5$years? 
Compared to Transneptunian Objects, Centaurs have different detectability because OSSOS detections are motion-rate dependent. With an upper limit for the motion rate set at 15"/hr, OSSOS necessarily focusses on objects detectable at more than 10~au, though it did detect objects inside that limit. In this regard, OSSOS may be biased toward more stable orbits found beyond Saturn \citep{TiscarenoMalhotra2003, DiSistoBrunini2007}, i.e. with perihelion distances larger than 10~au. More generally, we could argue that this stability bias is likely to be inherent in the Centaur discoveries by all surveys. There have been no surveys that adequately target the motion rate for objects in the 5-10~au region (where Centaurs with more unstable orbits can be found) in a well-characterized manner. The sensitivity of asteroid surveys drops beyond 5~au, and TNO surveys optimise cadence for objects typically beyond Saturn. So we could argue that this is a weakness of the entire known Centaur dataset, not just of OSSOS.

%==========================================================
\section{Absolute magnitudes and inferred sizes}
Because none of the OSSOS objects are measurably active, we can assume that the observed magnitude corresponds to the magnitude of the bare nucleus, and thus assess its size. At the discovery epoch OSSOS discovered moving objects in observations of each field with a triplet of exposures spaced over 2 hours. The quality of these discovery images was of crucial importance since the depth reached in these would set the limits for detecting objects. The photometric calibration of these images has been carefully done \citep[see][for the detailed procedure]{Bannister2018}, leading to $r-$band magnitudes (measured in an aperture of 5 times the Full Width at Half Maximum of the PSF) being extremely well characterized. 
We can convert them into absolute magnitudes $m_r(1,1,0)$, correcting for the heliocentric and geocentric distances ($R_{au}$ and $\Delta_{au}$ respectively) and the phase angle $\alpha$ of each detection (all paramaters given in Table \ref{size}), using the following equation:
\begin{equation}
m_r(1,1,0)=m_r - 5 log_{10}(R_{au}\Delta_{au}) - \Phi(\alpha)
\end{equation}

\begin{figure}
\includegraphics[width=\hsize]{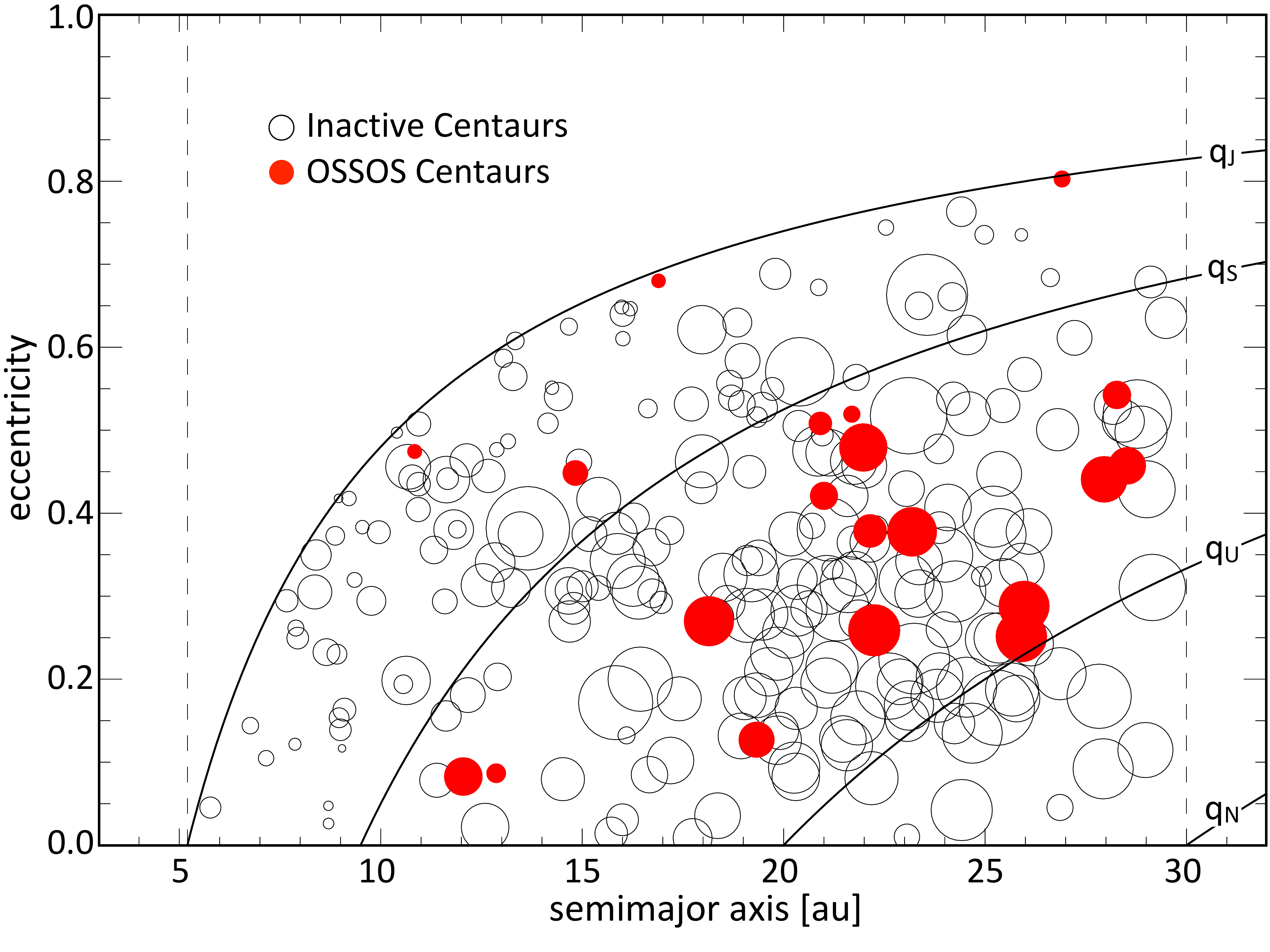}
      \caption{Distribution of Centaurs' absolute magnitude in the semi-major axis vs. eccentricity plane. Inactive Centaurs are shown in black (H values from the JPL Horizon database), OSSOS Centaurs are shown in red (computed in this work). The size of the symbols are proportional to the inverse of the absolute magnitude of Centaurs. Solid curves show the orbits having perihelia distances equal to the semi-major axis of Jupiter (q$_J$), Saturn (q$_S)$, Uranus (q$_U$) and Neptune (q$_N$).}
         \label{absolute}
   \end{figure}

For the phase function correction $\Phi(\alpha)$, we adopt $\Phi(\alpha) = 10^{-0.4\alpha\beta}$ with $\beta$=0.09~mag~deg$^{-1}$ for Centaurs \citep{Bauer2003}. The uncertainty on $m_R (1,1,0)$ has three components. The first comes from the uncertainty on $m_R$ itself, given in Table \ref{size}. The second comes from the phase function correction, which can be large and remains unknown. This problem has been discussed in detail by \citet{Jewitt2009}. The third comes from uncertainties on $R_{au}$ and $\Delta_{au}$: they are negligible for all objects except for the two $nt$ objects. The resulting absolute magnitudes are given in Table \ref{size} and displayed on Fig. \ref{absolute}, along with the JPL absolute magnitudes of inactive Centaurs.  

Computing sizes from our absolute magnitudes will necessarily be affected by the aforementioned uncertainties, and will only provide estimates. In addition, the conversion from absolute magnitude to size carries another source of uncertainty due to the albedo. Choosing the albedo for converting absolute magnitudes to a radius must be done with care. \cite{Bauer2013} found a mean albedo for Centaurs of $8\pm4\%$, consistent with the value from Duffard et al. (2014) of $7\pm5\%$. The later suggest that larger Centaurs have lower albedos, while smaller Centaurs span a wide range of albedos, typically between 4 and 16\%. No correlation between albedo and size, or albedo and orbital parameters can be found \citep{Duffard2014}. The only confirmed correlation is with their color. The bimodality of the colors of Centaurs was apparent very early \citep{Peixinho2003, Stansberry2008, Fraser2014}. \citet{Lacerda2014} showed that, in addition to having a bimodal distribution, colors are correlated with albedo for Centaurs and transneptunian objects in general. \citet{Bauer2013} and \citet{Duffard2014} come to the same conclusion, which has also been confirmed by \citet{Tegler2016} and \citet{RomanishinTegler2018}. In this framework neutral-gray objects are dark (albedo around 5\%), while red objects are bright (albedo around 15\%).

Since the only known information about the OSSOS Centaurs are their orbital parameters, which do not correlate with albedo, we choose to provide estimates for a range of albedos: 4\% and 16\% as limiting values, and 10\% as an average value. The resulting radii can be found in Table \ref{size}. The Col-OSSOS program provides \emph{g}-\emph{r} colors for 3 Centaurs in our sample (K14UM5J and K15RO5V have a \emph{g}-\emph{r} of 0.65$\pm$0.02 and 0.61$\pm$0.04 respectively \citep[][Schwamb et al. subm.\footnote{https://arxiv.org/abs/1809.08501}]{Pike2017}). These are blue objects, for which the albedo can be assumed to be low. Therefore, the preferred solution for their size would be the radii computed with the 4\% albedo. This also means that these Centaurs could have suffered from past cometary activity according to the analysis of small body populations' colors performed by \citet{Jewitt2015}.

\begin{table*}
\caption{Apparent and absolute magnitudes of OSSOS Centaurs, and estimated sizes for various albedos.\label{size}}
\centering
\begin{tabular}{lrrrcrcccrrrrrrr}
\hline\hline
\# \tablefootmark{a} & $R_{au}$ &$\Delta_{au}$ & $\alpha$ & $m_{lim}$\tablefootmark{b} & $\rho_{lim}$\tablefootmark{c} & $f_{lim}$ & $m_r$ & $\sigma_r$ & $m_r(1,1,0)$ & $\rho_{04}$\tablefootmark{d} & $\sigma_{04}$& $\rho_{10}$\tablefootmark{e} & $\sigma_{10}$ & $\rho_{16}$\tablefootmark{f} & $\sigma_{16}$ \\

 & [au] & [au] & [deg] & [mag] & [km] & [\%] &[mag] &  & [mag] & [km] & & [km] & &[km] & \\
 
 \hline
 
%o3o01 & 13.772 & 12.793 & 1.06 & 23.85 & 4.79 &  23.44 & 0.05 & 11.29 & 11.57 & 2.45 & 7.32 & 1.55 & 5.78 & 1.23 \\
1 & 13.772 & 12.793 & 1.06 & 23.85 & 4.79 & 63 & 23.35 & 0.07 & 11.20 & 12.06 & 3.01 & 7.62 & 1.91 & 6.03 & 1.51 \\
%           &             &             &         &           &         &  23.39 & 0.07 & 11.24 & 11.84 & 2.96 & 7.49 & 1.87 & 5.92 & 1.48 \\

%o3l02 & 17.045 & 16.131 & 1.34 & 24.42 & 5.81 &  24.17 & 0.16 & 11.08 & 13.05 & 4.83 & 8.25 & 3.05 & 6.52 & 2.41 \\
2 & 17.045 & 16.131 & 1.34 & 24.42 & 5.81 & 66 & 23.98 & 0.14 & 10.89 & 14.24 & 4.95 & 9.01 & 3.13 & 7.12 & 2.48 \\
%           &             &             &         &           &         &  23.57 & 0.09 & 10.48 & 17.21 & 4.85 & 10.88 & 3.07 & 8.60 & 2.43 \\

%o3l03 & 25.331 & 24.406 & 0.83 & 24.42 & 12.80 &  24.76 & 0.29 & 9.87 & 21.89 & 10.60 & 13.84 & 6.70 & 10.94 & 5.30 \\
3 & 25.331 & 24.406 & 0.83 & 24.42 & 12.80 & 68 & 24.01 & 0.13 & 9.12 & 30.92 & 10.39 & 19.56 & 6.57 & 15.46 & 5.19 \\
%           &             &             &         &           &         &  23.88 & 0.14 & 8.99 & 32.84 & 11.42 & 20.77 & 7.22 & 16.42 & 5.71 \\

%o4h01 & 17.751 & 16.758 & 0.23 & 24.55 & 5.66 &  22.97 & 0.06 & 9.62 & 23.43 & 5.43 & 14.82 & 3.44 & 11.71 & 2.72 \\
4 & 17.751 & 16.758 & 0.23 & 24.55 & 5.66 & 17 & 22.66 & 0.04 & 9.31 & 27.02 & 5.14 & 17.09 & 3.25 & 13.51 & 2.57 \\
%           &             &             &         &           &         &  22.58 & 0.03 & 9.23 & 28.04 & 4.63 & 17.73 & 2.93 & 14.02 & 2.31 \\
%           & 18.249 & 17.290 & 0.83 & 24.55 & 6.15 &  23.36 & 0.10 & 9.93 & 21.29 & 6.31 & 13.46 & 3.99 & 10.64 & 3.16 \\
%           &             &             &         &           &         &  23.08 & 0.07 & 9.65 & 24.22 & 6.05 & 15.32 & 3.83 & 12.11 & 3.03 \\
%           &             &             &         &           &         &  23.06 & 0.06 & 9.63 & 24.44 & 5.67 & 15.46 & 3.58 & 12.22 & 2.83 \\

%o4h02 & 19.520 & 18.526 & 0.17 & 24.55 & 6.86 &  24.45 & 0.21 & 10.67 & 14.37 & 6.03 & 9.09 & 3.81 & 7.19 & 3.01 \\
5 & 19.520 & 18.526 & 0.17 & 24.55 & 6.86 & 84 & 24.36 & 0.16 & 10.58 & 14.98 & 5.54 & 9.47 & 3.51 & 7.49 & 2.77 \\
%           &             &             &         &           &         &  24.30 & 0.15 & 10.52 & 15.40 & 5.53 & 9.74 & 3.50 & 7.70 & 2.77 \\

%o5p001 & 12.031 & 11.029 & 0.05 & 24.49 & 2.58 &  24.05 & 0.12 & 12.44 & 6.31 & 2.04 & 3.99 & 1.29 & 3.15 & 1.02 \\
6 & 12.031 & 11.029 & 0.05 & 24.49 & 2.58 & 59 & 23.92 & 0.12 & 12.31 & 6.70 & 2.17 & 4.24 & 1.37 & 3.35 & 1.08 \\
%           &             &             &         &           &         &  24.47 & 0.17 & 12.86 & 5.20 & 1.98 & 3.29 & 1.25 & 2.60 & 0.99 \\

%o5p002nt & 12.115 & 11.117 & 0.41 & 24.49 & 2.65 &  24.24 & 0.14 & 12.63 & 5.95 & 2.07 & 3.77 & 1.31 & 2.98 & 1.04 \\
7\tablefootmark{$\star$} & 12.115 & 11.117 & 0.41 & 24.49 & 2.65 & 68 & 24.08 & 0.14 & 12.47 & 6.41 & 2.23 & 4.05 & 1.41 & 3.20 & 1.11 \\
%           &             &             &         &           &         &  23.83 & 0.10 & 12.22 & 7.19 & 2.13 & 4.55 & 1.35 & 3.60 & 1.07 \\

%o5p003 & 13.565 & 12.564 & 0.22 & 24.49 & 3.33 &  21.34 & 0.01 & 9.20 & 28.42 & 2.72 & 17.98 & 1.72 & 14.21 & 1.36 \\
8 & 13.565 & 12.564 & 0.22 & 24.49 & 3.33 & 80 & 21.39 & 0.01 & 9.25 & 27.78 & 2.66 & 17.57 & 1.68 & 13.89 & 1.33 \\
%           &             &             &         &           &         &  21.45 & 0.01 & 9.31 & 27.02 & 2.59 & 17.09 & 1.64 & 13.51 & 1.29 \\
%           & 13.538 & 12.977 & 3.52 & 24.49 & 3.94 &  21.86 & 0.01 & 9.89 & 26.44 & 2.53 & 16.72 & 1.60 & 13.22 & 1.27 \\
%           &             &             &         &           &         &  21.86 & 0.01 & 9.89 & 26.44 & 2.53 & 16.72 & 1.60 & 13.22 & 1.27 \\
%           &             &             &         &           &         &  21.93 & 0.01 & 9.96 & 25.38 & 2.43 & 16.05 & 1.54 & 12.69 & 1.22 \\

%o5p004 & 13.564 & 12.563 & 0.22 & 24.49 & 3.33 &  23.29 & 0.06 & 11.15 & 11.58 & 2.68 & 7.32 & 1.70 & 5.79 & 1.34 \\
9 & 13.564 & 12.563 & 0.22 & 24.49 & 3.33 & 69 & 24.09 & 0.12 & 11.95 & 8.01 & 2.59 & 5.07 & 1.64 & 4.00 & 1.30 \\
%           &             &             &         &           &         &  23.75 & 0.08 & 11.61 & 9.36 & 2.50 & 5.92 & 1.58 & 4.68 & 1.25 \\

%o5p005 & 23.497 & 22.496 & 0.12 & 24.66 & 9.51 &  23.77 & 0.09 & 9.16 & 28.67 & 8.09 & 18.13 & 5.11 & 14.33 & 4.04 \\
10 & 23.497 & 22.496 & 0.12 & 24.66 & 9.51 & 40 & 23.66 & 0.08 & 9.05 & 30.16 & 8.04 & 19.07 & 5.08 & 15.08 & 4.02 \\
%           &             &             &         &           &         &  24.34 & 0.13 & 9.73 & 22.04 & 7.40 & 13.94 & 4.68 & 11.02 & 3.70 \\

%o5m01 & 7.434 & 6.432 & 1.21 & 24.42 & 1.01 &  23.60 & 0.07 & 14.30 & 2.93 & 0.73 & 1.86 & 0.46 & 1.47 & 0.37 \\
11 & 7.434 & 6.432 & 1.21 & 24.42 & 1.01 & 43 & 23.50 & 0.05 & 14.20 & 3.07 & 0.65 & 1.94 & 0.41 & 1.54 & 0.33 \\
%           &             &             &         &           &         &  23.56 & 0.05 & 14.26 & 2.99 & 0.63 & 1.89 & 0.40 & 1.49 & 0.32 \\

%o5m02 & 9.179 & 8.188 & 1.35 & 24.45 & 1.57 &  24.64 & 0.17 & 14.37 & 2.87 & 1.09 & 1.82 & 0.69 & 1.44 & 0.55 \\
12 & 9.179 & 8.188 & 1.35 & 24.45 & 1.57 & 88 & 24.32 & 0.11 & 14.05 & 3.33 & 1.03 & 2.11 & 0.65 & 1.67 & 0.52 \\
%           &             &             &         &           &         &  24.31 & 0.13 & 14.04 & 3.35 & 1.12 & 2.12 & 0.71 & 1.67 & 0.56 \\
%           & 9.160 & 8.155 & 0.76 & 24.45 & 1.52 &  24.31 & 0.14 & 14.00 & 3.24 & 1.13 & 2.05 & 0.71 & 1.62 & 0.56 \\
%           &             &             &         &           &         &  24.44 & 0.17 & 14.13 & 3.06 & 1.16 & 1.93 & 0.74 & 1.53 & 0.58 \\
%           &             &             &         &           &         &  24.38 & 0.17 & 14.07 & 3.14 & 1.20 & 1.99 & 0.76 & 1.57 & 0.60 \\

%o5s01 & 6.232 & 5.299 & 3.77 & 24.85 & 0.63 &  23.29 & 0.04 & 14.96 & 2.60 & 0.49 & 1.64 & 0.31 & 1.30 & 0.25 \\
13 & 6.232 & 5.299 & 3.77 & 24.85 & 0.63 & 27 & 23.43 & 0.04 & 15.10 & 2.44 & 0.46 & 1.54 & 0.29 & 1.22 & 0.23 \\
%           &             &             &         &           &         &  23.43 & 0.04 & 15.10 & 2.44 & 0.46 & 1.54 & 0.29 & 1.22 & 0.23 \\
%           & 6.141 & 5.146 & 0.98 & 24.85 & 0.54 &  22.64 & 0.06 & 14.22 & 2.99 & 0.69 & 1.89 & 0.44 & 1.49 & 0.35 \\
%           &             &             &         &           &         &  22.17 & 0.04 & 13.75 & 3.71 & 0.71 & 2.35 & 0.45 & 1.86 & 0.35 \\
%           &             &             &         &           &         &  22.26 & 0.04 & 13.84 & 3.56 & 0.68 & 2.25 & 0.43 & 1.78 & 0.34 \\

%o5s03nt & 11.747 & 10.801 & 1.76 & 24.85 & 2.24 &  24.80 & 0.12 & 13.42 & 4.58 & 1.48 & 2.90 & 0.94 & 2.29 & 0.74 \\
14\tablefootmark{$\star$} & 11.747 & 10.801 & 1.76 & 24.85 & 2.24 & 83 & 24.65 & 0.10 & 13.27 & 4.91 & 1.46 & 3.11 & 0.92 & 2.46 & 0.73 \\
%           &             &             &         &           &         &  24.84 & 0.12 & 13.46 & 4.50 & 1.46 & 2.84 & 0.92 & 2.25 & 0.73 \\

%o5s04 & 13.434 & 12.494 & 1.60 & 25.23 & 2.47 &  24.51 & 0.09 & 12.51 & 6.88 & 1.94 & 4.35 & 1.23 & 3.44 & 0.97 \\
15 & 13.434 & 12.494 & 1.60 & 25.23 & 2.47 & 58 & 24.64 & 0.09 & 12.64 & 6.48 & 1.83 & 4.10 & 1.16 & 3.24 & 0.91 \\
%           &             &             &         &           &         &  24.56 & 0.09 & 12.56 & 6.72 & 1.90 & 4.25 & 1.20 & 3.36 & 0.95 \\

%o5s05 & 19.874 & 18.948 & 1.16 & 25.15 & 5.65 &  23.21 & 0.04 & 9.42 & 27.60 & 5.25 & 17.45 & 3.32 & 13.80 & 2.62 \\
16 & 19.874 & 18.948 & 1.16 & 25.15 & 5.65 & 17 & 23.25 & 0.04 & 9.46 & 27.09 & 5.15 & 17.13 & 3.26 & 13.55 & 2.58 \\
%           &             &             &         &           &         &  23.21 & 0.04 & 9.42 & 27.60 & 5.25 & 17.45 & 3.32 & 13.80 & 2.62 \\
%           & 19.823 & 19.162 & 2.26 & 25.15 & 5.96 &  23.61 & 0.09 & 9.88 & 24.23 & 6.84 & 15.33 & 4.32 & 12.12 & 3.42 \\
%           &             &             &         &           &         &  23.05 & 0.09 & 9.32 & 31.35 & 8.84 & 19.83 & 5.59 & 15.67 & 4.42 \\
%           &             &             &         &           &         &  23.12 & 0.09 & 9.39 & 30.36 & 8.56 & 19.20 & 5.41 & 15.18 & 4.28 \\
%           & 19.835 & 19.099 & 2.05 & 25.15 & 5.89 &  23.17 & 0.08 & 9.43 & 29.34 & 7.82 & 18.56 & 4.95 & 14.67 & 3.91 \\
%           &             &             &         &           &         &  23.79 & 0.12 & 10.05 & 22.04 & 7.13 & 13.94 & 4.51 & 11.02 & 3.57 \\
%           &             &             &         &           &         &  22.99 & 0.07 & 9.25 & 31.86 & 7.96 & 20.15 & 5.04 & 15.93 & 3.98 \\

%o5t02 & 10.612 & 9.683 & 2.20 & 24.68 & 2.00 &  24.47 & 0.15 & 13.58 & 4.40 & 1.58 & 2.78 & 1.00 & 2.20 & 0.79 \\
17 & 10.612 & 9.683 & 2.20 & 24.68 & 2.00 & 92 & 24.60 & 0.17 & 13.71 & 4.14 & 1.58 & 2.62 & 1.00 & 2.07 & 0.79 \\
%           &             &             &         &           &         &  24.91 & 0.19 & 14.02 & 3.59 & 1.44 & 2.27 & 0.91 & 1.80 & 0.72 \\

%o5t03 & 18.511 & 17.575 & 1.18 & 24.97 & 5.30 &  23.16 & 0.04 & 9.69 & 24.42 & 4.64 & 15.44 & 2.94 & 12.21 & 2.32 \\
18 & 18.511 & 17.575 & 1.18 & 24.97 & 5.30 & 21 & 23.27 & 0.04 & 9.80 & 23.21 & 4.41 & 14.68 & 2.79 & 11.61 & 2.21 \\
%           &             &             &         &           &         &  23.26 & 0.04 & 9.79 & 23.32 & 4.43 & 14.75 & 2.80 & 11.66 & 2.22 \\
%           & 18.515 & 17.522 & 0.39 & 24.97 & 5.12 &  22.35 & 0.04 & 8.83 & 34.22 & 6.51 & 21.64 & 4.12 & 17.11 & 3.25 \\
%           &             &             &         &           &         &  22.70 & 0.06 & 9.18 & 29.13 & 6.75 & 18.42 & 4.27 & 14.56 & 3.38 \\
%           &             &             &         &           &         &  22.39 & 0.04 & 8.87 & 33.60 & 6.39 & 21.25 & 4.04 & 16.80 & 3.19 \\

%o5c001 & 15.853 & 14.866 & 0.35 & 24.67 & 4.26 &  23.90 & 0.08 & 11.07 & 12.15 & 3.24 & 7.69 & 2.05 & 6.08 & 1.62 \\
19 & 15.853 & 14.866 & 0.35 & 24.67 & 4.26 & 35 & 23.54 & 0.06 & 10.71 & 14.35 & 3.33 & 9.07 & 2.10 & 7.17 & 1.66 \\
%           &             &             &         &           &         &  23.84 & 0.08 & 11.01 & 12.49 & 3.33 & 7.90 & 2.11 & 6.25 & 1.66 \\

%o5d001 & 13.282 & 12.296 & 0.43 & 24.82 & 2.77 &  24.00 & 0.06 & 11.97 & 8.07 & 1.87 & 5.10 & 1.18 & 4.04 & 0.94 \\
20 & 13.282 & 12.296 & 0.43 & 24.82 & 2.77 & 43 & 23.94 & 0.07 & 11.91 & 8.30 & 2.07 & 5.25 & 1.31 & 4.15 & 1.04 \\
%           &             &             &         &           &         &  23.92 & 0.06 & 11.89 & 8.37 & 1.94 & 5.30 & 1.23 & 4.19 & 0.97 \\
\end{tabular}
\tablefoot{
\tablefoottext{a}{See corresponding name with Table \ref{orbparam}.}
\tablefoottext{b}{Detection limit set in magnitude, from \citet{Bannister2018} and accounting for the object's motion rate.}
\tablefoottext{c}{Radius corresponding to the detection limit assuming a 16\% albedo, i.e. radius of the smallest object detectable.}
\tablefoottext{d}{Radius of the object assuming a 4\% albedo.}
\tablefoottext{e}{Radius of the object assuming a 10\% albedo.}
\tablefoottext{f}{Radius of the object assuming a 16\% albedo.}
\tablefoottext{$\star$}{We stress again that these objects were not tracked and have intrinsic uncertainties much larger than given in the table, though they cannot be quantified.}
}
\end{table*}

%==========================================================
\section{Discussion}
With 20 objects discovered in the giant planet region by OSSOS, we could have expected a few of them to be active, especially for Centaurs such as K15KH2H, K15KH2J and K15RR7K
%o5m01, o5m02 and o5s01 
with perihelion distances in the 5-6 au range. Our search for cometary activity has revealed that no coma can be detected in the OSSOS dataset.
However, because the search for activity is done at the detection level of objects themselves, we are sensitive to high levels of activity for most of these objects, and therefore cannot rule out that they are active below our detection limits. The deeper Col-OSSOS dataset available for 3 objects does not reveal any activity either.
While active Centaurs tend to have perihelion distances in the Jupiter-Saturn range, most Centaurs in this region remain inactive despite having small perihelia. In addition, \citet{Jewitt2009} showed that the lack of active Centaurs further away from the Sun is not the consequence of an observational bias, but rather the manifestation of the underlying process driving the activity of Centaurs. They proposed that the crystallization of amorphous water ice, releasing trapped volatiles, would be able to fit the overall dataset on active and inactive Centaurs. 
Other sources of activity, in particular the sublimation of ices more volatile than water such as CO$_2$, CO, N$_2$ or O$_2$, remain possible, as recently observed for long-period comet C/2017 K2 (PANSTARRS) \citep{Jewitt2017}. However, they would require that active Centaurs should be observed at all heliocentric distances \citep{Jewitt2009}, which is not the case. In addition, no strong detection of gaseous CO in a Centaur other than 29P/Schwassmann-Wachmann 1 has been reported to date \citep{SenayJewitt1994, Crovisier1995, Gunnarsson2008}. The very deep search for the J(2-1) rotational line of CO performed by \citet{Drahus2017} provides the most sensitive CO production rates to date, and are consistent with the activity of Centaurs not being driven by the sublimation of CO.

Crystallization as the source for cometary activity amongst Centaurs was studied by \citet{GuilbertLepoutre2012}. They showed that the phase transition could sustain cometary activity up to 10-12 au (in equivalent semimajor axis defined by a$_{eq}$=$a(1-e^2)$) albeit for a limited time. This corresponds to the timescale for the crystallization front to propagate under the surface at a depth where molecules released by the phase transition cannot escape the object anymore, and thus cannot contribute to any significant outgassing. The direct consequence of this effect is that in order to be active, a Centaur needs to have suffered from a recent drop in its perihelion distance, or rather in its equivalent semimajor axis. In addition, even if such a drop occurs, triggering the activity would require that the object did not previously stay in the "active" region for a long period of time, or its near-surface layers would be depleted in volatiles or fully crystallized already as a result of past activity (as supported by \citealt{Fernandez2018}). 

The lack of activity in the OSSOS dataset does no refute this scenario. We can even argue that the orbital evolution of OSSOS Centaurs is consistent with it, since none fits the thermal requirements for triggering crystallization-driven activity. At this stage, we should stress again that these objects may be active below or detection limits, that there is still no direct proof that amorphous water ice is present in Centaurs and that crystallization is driving their activity. Another phase transition, occurring in the giant planet region up to 10-12 au, limited in time and irreversible, could also explain the dataset. To our knowledge, crystallization of amorphous water ice is -- as of today -- the only phase transition which fits these constraints and provide a plausible source of outgassing for Centaurs. We agree with \citet{Fernandez2018} that most active and inactive Centaurs differ only in their individual dynamical evolutions. We further argue that if crystallization is the source of cometary activity amongst Centaurs, it would favor activity on objects dynamically new to the Jupiter-Saturn region, as objects having previously stayed in this region would have exhausted their amorphous water ice in the near-surface layers.

%==========================================================
\section{Conclusion}
We searched for cometary activity among the new Centaurs detected by the Outer Solar System Origins Survey (OSSOS). 
Our results can be summarized as follows:
\begin{enumerate}
\item Though some objects are close enough to the Sun to be potentially active, no coma can be detected either in the OSSOS dataset, or the Col-OSSOS data when available.
\item The analysis of their past orbital evolution shows that none of the OSSOS Centaurs meets the thermal requirements for being active due to the crystallization of amorphous water ice.
\item The properties of OSSOS Centaurs support (or at least do not provide any evidence against) the current scenario in which crystallization of amorphous water ice is the source for cometary activity amongst Centaurs. 
\end{enumerate}

\begin{acknowledgements}
We are grateful to Colin Snodgrass for comments and suggestions which improved the manuscript.
\end{acknowledgements}

\bibliographystyle{aa}
\bibliography{Reference}

\begin{thebibliography}{43}
\expandafter\ifx\csname natexlab\endcsname\relax\def\natexlab#1{#1}\fi

\bibitem[{{Bannister} {et~al.}(2018){Bannister}, {Gladman}, {Kavelaars},
  {Petit}, {Volk}, {Chen}, {Alexandersen}, {Gwyn}, {Schwamb}, {Ashton},
  {Benecchi}, {Cabral}, {Dawson}, {Delsanti}, {Fraser}, {Granvik},
  {Greenstreet}, {Guilbert-Lepoutre}, {Ip}, {Jakubik}, {Jones}, {Kaib},
  {Lacerda}, {Van Laerhoven}, {Lawler}, {Lehner}, {Lin}, {Lykawka}, {Marsset},
  {Murray-Clay}, {Pike}, {Rousselot}, {Shankman}, {Thirouin}, {Vernazza}, \&
  {Wang}}]{Bannister2018}
{Bannister}, M.~T., {Gladman}, B.~J., {Kavelaars}, J.~J., {et~al.} 2018, \apjs,
  236, 18

\bibitem[{{Bannister} {et~al.}(2016){Bannister}, {Kavelaars}, {Petit},
  {Gladman}, {Gwyn}, {Chen}, {Volk}, {Alexandersen}, {Benecchi}, {Delsanti},
  {Fraser}, {Granvik}, {Grundy}, {Guilbert-Lepoutre}, {Hestroffer}, {Ip},
  {Jakubik}, {Jones}, {Kaib}, {Kavelaars}, {Lacerda}, {Lawler}, {Lehner},
  {Lin}, {Lister}, {Lykawka}, {Monty}, {Marsset}, {Murray-Clay}, {Noll},
  {Parker}, {Pike}, {Rousselot}, {Rusk}, {Schwamb}, {Shankman}, {Sicardy},
  {Vernazza}, \& {Wang}}]{Bannister2016}
{Bannister}, M.~T., {Kavelaars}, J.~J., {Petit}, J.-M., {et~al.} 2016, \aj,
  152, 70

\bibitem[{{Bar-Nun} {et~al.}(1985){Bar-Nun}, {Herman}, {Laufer}, \&
  {Rappaport}}]{BarNun1985}
{Bar-Nun}, A., {Herman}, G., {Laufer}, D., \& {Rappaport}, M.~L. 1985, \icarus,
  63, 317

\bibitem[{{Bar-Nun} \& {Owen}(1998)}]{BarNunOwen1998}
{Bar-Nun}, A. \& {Owen}, T. 1998, in Astrophysics and Space Science Library,
  Vol. 227, Solar System Ices, ed. B.~{Schmitt}, C.~{de Bergh}, \& M.~{Festou},
  353

\bibitem[{{Bauer} {et~al.}(2013){Bauer}, {Grav}, {Blauvelt}, {Mainzer},
  {Masiero}, {Stevenson}, {Kramer}, {Fern{\'a}ndez}, {Lisse}, {Cutri},
  {Weissman}, {Dailey}, {Masci}, {Walker}, {Waszczak}, {Nugent}, {Meech},
  {Lucas}, {Pearman}, {Wilkins}, {Watkins}, {Kulkarni}, {Wright}, {WISE Team},
  \& {PTF Team}}]{Bauer2013}
{Bauer}, J.~M., {Grav}, T., {Blauvelt}, E., {et~al.} 2013, \apj, 773, 22

\bibitem[{{Bauer} {et~al.}(2003){Bauer}, {Meech}, {Fern{\'a}ndez},
  {Pittichova}, {Hainaut}, {Boehnhardt}, \& {Delsanti}}]{Bauer2003}
{Bauer}, J.~M., {Meech}, K.~J., {Fern{\'a}ndez}, Y.~R., {et~al.} 2003, \icarus,
  166, 195

\bibitem[{{Bertin} \& {Arnouts}(1996)}]{BertinArnouts1996}
{Bertin}, E. \& {Arnouts}, S. 1996, \aaps, 117, 393

\bibitem[{{Crovisier} {et~al.}(1995){Crovisier}, {Biver}, {Bockelee-Morvan},
  {Colom}, {Jorda}, {Lellouch}, {Paubert}, \& {Despois}}]{Crovisier1995}
{Crovisier}, J., {Biver}, N., {Bockelee-Morvan}, D., {et~al.} 1995, \icarus,
  115, 213

\bibitem[{{Di Sisto} \& {Brunini}(2007)}]{DiSistoBrunini2007}
{Di Sisto}, R.~P. \& {Brunini}, A. 2007, \icarus, 190, 224

\bibitem[{{di Sisto} {et~al.}(2010){di Sisto}, {Brunini}, \& {de
  El{\'{\i}}a}}]{DiSisto2010}
{di Sisto}, R.~P., {Brunini}, A., \& {de El{\'{\i}}a}, G.~C. 2010, \aap, 519,
  A112

\bibitem[{{Drahus} {et~al.}(2017){Drahus}, {Yang}, {Lis}, \&
  {Jewitt}}]{Drahus2017}
{Drahus}, M., {Yang}, B., {Lis}, D.~C., \& {Jewitt}, D. 2017, \mnras, 468, 2897

\bibitem[{{Duffard} {et~al.}(2014){Duffard}, {Pinilla-Alonso}, {Santos-Sanz},
  {Vilenius}, {Ortiz}, {Mueller}, {Fornasier}, {Lellouch}, {Mommert}, {Pal},
  {Kiss}, {Mueller}, {Stansberry}, {Delsanti}, {Peixinho}, \&
  {Trilling}}]{Duffard2014}
{Duffard}, R., {Pinilla-Alonso}, N., {Santos-Sanz}, P., {et~al.} 2014, \aap,
  564, A92

\bibitem[{{Duncan} \& {Levison}(1997)}]{DuncanLevison1997}
{Duncan}, M.~J. \& {Levison}, H.~F. 1997, Science, 276, 1670

\bibitem[{{Fern{\'a}ndez} {et~al.}(2018){Fern{\'a}ndez}, {Helal}, \&
  {Gallardo}}]{Fernandez2018}
{Fern{\'a}ndez}, J.~A., {Helal}, M., \& {Gallardo}, T. 2018, \planss, 158, 6

\bibitem[{{Fraser} {et~al.}(2016){Fraser}, {Alexandersen}, {Schwamb},
  {Marsset}, {Pike}, {Kavelaars}, {Bannister}, {Benecchi}, \&
  {Delsanti}}]{Fraser2016}
{Fraser}, W., {Alexandersen}, M., {Schwamb}, M.~E., {et~al.} 2016, \aj, 151,
  158

\bibitem[{{Fraser} {et~al.}(2014){Fraser}, {Brown}, {Morbidelli}, {Parker}, \&
  {Batygin}}]{Fraser2014}
{Fraser}, W.~C., {Brown}, M.~E., {Morbidelli}, A., {Parker}, A., \& {Batygin},
  K. 2014, \apj, 782, 100

\bibitem[{{Guilbert-Lepoutre}(2012)}]{GuilbertLepoutre2012}
{Guilbert-Lepoutre}, A. 2012, \aj, 144, 97

\bibitem[{{Gunnarsson} {et~al.}(2008){Gunnarsson}, {Bockel{\'e}e-Morvan},
  {Biver}, {Crovisier}, \& {Rickman}}]{Gunnarsson2008}
{Gunnarsson}, M., {Bockel{\'e}e-Morvan}, D., {Biver}, N., {Crovisier}, J., \&
  {Rickman}, H. 2008, \aap, 484, 537

\bibitem[{{Hodapp} {et~al.}(2003){Hodapp}, {Jensen}, {Irwin}, {Yamada},
  {Chung}, {Fletcher}, {Robertson}, {Hora}, {Simons}, {Mays}, {Nolan}, {Bec},
  {Merrill}, \& {Fowler}}]{Hodapp}
{Hodapp}, K.~W., {Jensen}, J.~B., {Irwin}, E.~M., {et~al.} 2003, \pasp, 115,
  1388

\bibitem[{{Hook} {et~al.}(2004){Hook}, {J{\o}rgensen}, {Allington-Smith},
  {Davies}, {Metcalfe}, {Murowinski}, \& {Crampton}}]{Hook}
{Hook}, I.~M., {J{\o}rgensen}, I., {Allington-Smith}, J.~R., {et~al.} 2004,
  \pasp, 116, 425

\bibitem[{{Horner} {et~al.}(2004){Horner}, {Evans}, \& {Bailey}}]{Horner2004}
{Horner}, J., {Evans}, N.~W., \& {Bailey}, M.~E. 2004, \mnras, 354, 798

\bibitem[{{Horner} \& {Lykawka}(2010)}]{HornerLykawka2010}
{Horner}, J. \& {Lykawka}, P.~S. 2010, \mnras, 402, 13

\bibitem[{{Hudson} \& {Donn}(1991)}]{HudsonDonn1991}
{Hudson}, R.~L. \& {Donn}, B. 1991, \icarus, 94, 326

\bibitem[{{Jenniskens} \& {Blake}(1994)}]{JenniskensBlake1994}
{Jenniskens}, P. \& {Blake}, D.~F. 1994, Science, 265, 753

\bibitem[{{Jewitt}(2009)}]{Jewitt2009}
{Jewitt}, D. 2009, \aj, 137, 4296

\bibitem[{{Jewitt}(2015)}]{Jewitt2015}
{Jewitt}, D. 2015, \aj, 150, 201

\bibitem[{{Jewitt} {et~al.}(2017){Jewitt}, {Hui}, {Mutchler}, {Weaver}, {Li},
  \& {Agarwal}}]{Jewitt2017}
{Jewitt}, D., {Hui}, M.-T., {Mutchler}, M., {et~al.} 2017, \apjl, 847, L19

\bibitem[{{Lacerda} {et~al.}(2014){Lacerda}, {Fornasier}, {Lellouch}, {Kiss},
  {Vilenius}, {Santos-Sanz}, {Rengel}, {M{\"u}ller}, {Stansberry}, {Duffard},
  {Delsanti}, \& {Guilbert-Lepoutre}}]{Lacerda2014}
{Lacerda}, P., {Fornasier}, S., {Lellouch}, E., {et~al.} 2014, \apjl, 793, L2

\bibitem[{{Laufer} {et~al.}(1987){Laufer}, {Kochavi}, \&
  {Bar-Nun}}]{Laufer1987}
{Laufer}, D., {Kochavi}, E., \& {Bar-Nun}, A. 1987, \prb, 36, 9219

\bibitem[{Lawler {et~al.}(2018)Lawler, Kavelaars, Alexandersen, Bannister,
  Gladman, Petit, \& Shankman}]{Lawler2018}
Lawler, S.~M., Kavelaars, J.~J., Alexandersen, M., {et~al.} 2018, Frontiers in
  Astronomy and Space Sciences, 5, 14

\bibitem[{{Levison} \& {Duncan}(1994)}]{LevisonDuncan1994}
{Levison}, H.~F. \& {Duncan}, M.~J. 1994, \icarus, 108, 18

\bibitem[{{Levison} \& {Duncan}(1997)}]{LevisonDuncan1997}
{Levison}, H.~F. \& {Duncan}, M.~J. 1997, \icarus, 127, 13

\bibitem[{{Notesco} \& {Bar-Nun}(1996)}]{NotescoBarNun1996}
{Notesco}, G. \& {Bar-Nun}, A. 1996, \icarus, 122, 118

\bibitem[{{Notesco} {et~al.}(2003){Notesco}, {Bar-Nun}, \&
  {Owen}}]{Notesco2003}
{Notesco}, G., {Bar-Nun}, A., \& {Owen}, T. 2003, \icarus, 162, 183

\bibitem[{{Peixinho} {et~al.}(2003){Peixinho}, {Doressoundiram}, {Delsanti},
  {Boehnhardt}, {Barucci}, \& {Belskaya}}]{Peixinho2003}
{Peixinho}, N., {Doressoundiram}, A., {Delsanti}, A., {et~al.} 2003, \aap, 410,
  L29

\bibitem[{{Pike} {et~al.}(2017){Pike}, {Fraser}, {Schwamb}, {Kavelaars},
  {Marsset}, {Bannister}, {Lehner}, {Wang}, {Alexandersen}, {Chen}, {Gladman},
  {Gwyn}, {Petit}, \& {Volk}}]{Pike2017}
{Pike}, R.~E., {Fraser}, W.~C., {Schwamb}, M.~E., {et~al.} 2017, \aj, 154, 101

\bibitem[{{Romanishin} \& {Tegler}(2018)}]{RomanishinTegler2018}
{Romanishin}, W. \& {Tegler}, S.~C. 2018, \aj, 156, 19

\bibitem[{{Senay} \& {Jewitt}(1994)}]{SenayJewitt1994}
{Senay}, M.~C. \& {Jewitt}, D. 1994, \nat, 371, 229

\bibitem[{{Snodgrass} {et~al.}(2016){Snodgrass}, {Jehin}, {Manfroid}, {Opitom},
  {Fitzsimmons}, {Tozzi}, {Faggi}, {Yang}, {Knight}, {Conn}, {Lister},
  {Hainaut}, {Bramich}, {Lowry}, {Rozek}, {Tubiana}, \&
  {Guilbert-Lepoutre}}]{Snodgrass2016}
{Snodgrass}, C., {Jehin}, E., {Manfroid}, J., {et~al.} 2016, \aap, 588, A80

\bibitem[{{Stansberry} {et~al.}(2008){Stansberry}, {Grundy}, {Brown},
  {Cruikshank}, {Spencer}, {Trilling}, \& {Margot}}]{Stansberry2008}
{Stansberry}, J., {Grundy}, W., {Brown}, M., {et~al.} 2008, {Physical
  Properties of Kuiper Belt and Centaur Objects: Constraints from the Spitzer
  Space Telescope}, ed. M.~A. {Barucci}, H.~{Boehnhardt}, D.~P. {Cruikshank},
  A.~{Morbidelli}, \& R.~{Dotson}, 161--179

\bibitem[{{Tegler} {et~al.}(2016){Tegler}, {Romanishin}, {Consolmagno}, \&
  {J.}}]{Tegler2016}
{Tegler}, S.~C., {Romanishin}, W., {Consolmagno}, G.~J., \& {J.}, S. 2016, \aj,
  152, 210

\bibitem[{{Tiscareno} \& {Malhotra}(2003)}]{TiscarenoMalhotra2003}
{Tiscareno}, M.~S. \& {Malhotra}, R. 2003, \aj, 126, 3122

\bibitem[{{Volk} \& {Malhotra}(2008)}]{VolkMalhotra2008}
{Volk}, K. \& {Malhotra}, R. 2008, \apj, 687, 714

\end{thebibliography}

\end{document}